\documentclass[prl,preprint,showpacs,preprintnumbers,amsmath,amssymb]{revtex4} %twocolumn

\usepackage{graphicx}% Include figure files
\usepackage{dcolumn}% Align table columns on decimal point
\usepackage{bm}% bold math
\begin{document}
\title{Network-Induced Oscillatory Behavior in Material Flow Networks} 
\author{Dirk Helbing} 
%  \email{helbing@trafficforum.org}
%  \homepage{http://www.helbing.org}
\affiliation{Dresden University of Technology, Andreas-Schubert-Str. 23, 
01069 Dresden, Germany}

\author{Ulrich Witt}
%\email{witt@mpiew-jena.mpg.de}
\affiliation{Max Planck Institute for Research into Economic Systems,
Kahlaische Stra{\ss}e 10, D-07745 Jena, Germany}

\author{Stefan L\"ammer}
\affiliation{Dresden University of Technology, Andreas-Schubert-Str. 23, 
01069 Dresden, Germany}

\author{Thomas Brenner}
\affiliation{Max Planck Institute for Research into Economic Systems,
Kahlaische Stra{\ss}e 10, D-07745 Jena, Germany}

\date{\today}

%47.54.+r Pattern selection; pattern formation
%89.65.Gh Economics, business, and financial markets
%89.40.+k Transportation
%89.20.Bb Industrial and technological research and development
%%47.62.+q Flow control
%47.55.-t Nonhomogeneous flows
%47.70.-n Reactive, radiative, or nonequilibrium flows
%%89.20.-a Interdisciplinary applications of physics
%%89.75.-k Complex systems
%%89.75.Da Systems obeying scaling laws
%%89.75.Fb Structures and organization in complex systems
%%89.75.Hc Networks and genealogical trees
%%89.75.Kd Patterns
%89.65.Gh Economics; econophysics, financial markets, business and
%management
%89.60.Gg Impact of natural and man-made disasters
%07.05.Kf Data analysis: algorithms and implementation; data management

\begin{abstract}
Network theory is rapidly changing our understanding of complex systems, but the relevance
of topological features for the \textit{dynamic} behavior of metabolic networks, food webs, production systems, 
information networks, or cascade failures of power grids remains to be explored. Based on a simple
model of supply networks, we offer an interpretation of instabilities and oscillations observed in 
biological, ecological, economic, and engineering systems. We find that most supply networks 
display \textit{damped oscillations,} even when their 
units---and linear chains of these units---behave in a \textit{non-oscillatory} way. 
Moreover, networks of \textit{damped} oscillators tend to produce \textit{growing} oscillations. 
This surprising behavior offers, for example, a new interpretation of business cycles
and of oscillating or pulsating processes.
The network structure of material flows itself turns out to be a source of instability,
and cyclical variations are an inherent feature of decentralized adjustments. %in delivery or production rates.
\end{abstract}
\pacs{89.75.Hc,%Networks and genealogical trees
89.65.Gh,%Economics; econophysics, financial markets, business and management
47.70.-n,%Reactive, radiative, or nonequilibrium flows
89.40.-a}%Transportation
\maketitle
\draft
Network theory \cite{strogatz,nets} is an important key to explaining complex systems. 
Topological features \cite{motifs} determine interesting
properties of socio-economic networks \cite{socio} or the robustness
of the world wide web \cite{www}, of sensory systems \cite{Alon}, and food webs \cite{Neutel}.
Their relevance for dynamic features, however, is not well understood.
Specific network structures appear to be responsible for the beating of a
leech's heart \cite{Stewart}, pumping effects in some slime molds \cite{Ueda}, strong variations 
in ecosystems \cite{Eco}, or glycolytic oscillations in yeast cells \cite{Dane}. 
Many of those systems, including molecular networks \cite{metabo}, 
can be viewed as particular supply networks. Their functioning
is decisive for natural and man-made systems, but prone
to oscillations: For example, industries often suffer from over-reactions 
in production \cite{supp}, disaster management struggles with a temporary clustering of forces and materials at some
places, while they are missing at others \cite{disaster}, and most economies are characterized by ``booms''
and ``recessions'' \cite{Mulli,econo}. 
Based on a simple model of interactive flows, we will show that the likely reason for these 
instabilities and oscillations is the underlying network structure of supply systems. 
This is illustrated with a model of macroscopic commodity flows.
\par
Our simple model of supply networks describes the generation or delivery of 
products, substances, materials, or other resources of kind $i$ at a certain rate $Q_i(t)\ge 0$
as a function of time $t$. A share $X_{ij}(t)\ge 0$ of that output is 
required as input for generating or delivering resource $j$, and 
a share $Y_i(t)\ge 0$ is absorbed, for example by consumption, degeneration, or loss.
If $\sum_j X_{ij}(t) + Y_i(t)$ does not add up to $Q_i(t)$, this 
is assumed to result in a corresponding change\linebreak $[N_i(t+\Delta t) - N_i(t)]$
of the stock, inventory, or concentration $N_i(t)$ of resource $i$ per unit time $\Delta t$.
This assumption implies a \textit{conservation equation} for resources (which can be generalized). 
In the limit $\Delta t \rightarrow 0$ it reads
\begin{equation}
 \frac{dN_i(t)}{dt} = \underbrace{Q_i(t)}_{\rm supply} - 
 \underbrace{\bigg[\sum_{j=1}^m a_{ij} Q_j(t) + \overbrace{Y_i(t)}^{\rm outflow}\!\bigg]}_{\rm demand} \, ,
\label{N}
\end{equation}
considering that the flow $X_{ij}(t)$ of resources from supplier $i$ 
represents a share $a_{ij}Q_j(t)\ge 0$ of the entire inputs used by supplier $j$. 
In Leontief's input-output model of macroeconomics \cite{Leontief}, the $m$ suppliers correspond to
different economic sectors producing one kind of commodity $i$ each. 
$Q_i(t)$ describes the output flow and $N_i(t)$ the inventory
of commodity $i$. The stoichiometric
coefficients $a_{ij} \le 1$ define an input matrix ${\mathbf{A}}=(a_{ij})$
and reflect the technologically determined supply network between economic sectors $i$ and $j$. 
For simplicity, we will treat the input coefficients $a_{ij}$ as constants. 
\par
The change of the delivery rate $Q_i(t)$ in time is generally a function $W_i$ of the stock levels or concentrations
$N_j$, their temporal change $dN_j/dt$, and the delivery rates $Q_j$ themselves:
$ dQ_i/dt = W_i(\{N_j\},\{dN_j/dt\},\{Q_j\})$ with $j \in \{1,\dots,m\}$.
In our model of macroeconomic output flows, the delivery rates $Q_i(t)$ are adjusted in response to two criteria:  
First, if the current inventory $N_i(t)$ exceeds some ``optimal'' level $N_i^0$, 
the delivery rate is reduced and vice versa. A certain inventory level is desireable to cope 
with variations in the demand. Second, if inventories are growing  
($dN_i/dt > 0$), i.e. if the current supply $Q_i(t)$ exceeds the current demand 
$Y_i(t) + \sum_j a_{ij}Q_j(t)$, this is an independent reason to reduce the delivery rate $Q_i(t)$.
The following equation for the relative change in the delivery rate expresses these responses 
and guarantees a non-negativity of $Q_i(t)$:
\begin{equation}
  \frac{1}{Q_i(t)} \frac{dQ_i}{dt} = \hat{\nu}_i 
\left( \frac{N_i^0}{N_i(t)} - 1 \right) - \frac{\hat{\mu}_i}{N_i(t)}  
\frac{dN_i}{dt} \, . 
\label{Q}
\end{equation}
Here, $\hat{\nu}_i$ is an adaptation rate describing the sensitivity to deviations of the actual 
inventory $N_i(t)$ from the desired one $N_i^0$. $\hat{\mu}_i$ is a dimensionless parameter
reflecting the responsiveness to relative deviations $(dN_i/dt)/N_i(t)$ from the stationary
equilibrium state. Moreover, economic systems have an important additional equilibration mechanism,
as undesired inventory levels and inventory changes can be compensated for by adjusting 
the price $P_i(t)$ of commodity $i$. If the same criteria are applied, we find 
\begin{equation}
  \frac{1}{P_i(t)} \frac{dP_i}{dt} = \nu_i \left( \frac{N_i^0}{N_i(t)} - 1 \right) -
 \frac{\mu_i}{N_i(t)} \frac{dN_i}{dt} \, . 
\label{P}
\end{equation}
We will assume that $\alpha_i = \hat{\nu}_i/\nu_i$ is the ratio between the adjustment 
rate of the output flow and the adjustment rate of the price in sector $i$. For simplicity,
the same ratio $\alpha_i = \hat{\mu}_i/\mu_i$ is assumed for the responsiveness.
\par
Increased prices $P_i(t)$ have a negative impact on the consumption rate $Y_i(t)$
and vice versa, which is described here by a standard demand function $f_i$ with
a negative derivative $f'_i(P_i)= df_i(P_i)/dP_i$:
\begin{equation}
 Y_i(t) = [Y_i^0 + \xi_i(t)] f_i(P_i(t)) \, . 
\label{Y}
\end{equation}
This formula takes into account 
random fluctuations $\xi_i(t)$ over time around a certain average consumption rate $Y_i^0$ and assumes 
that the average value of $f_i(P_i(t))$ is normalized to one. 
\par
The possible dynamic behaviors of supply networks can be studied by analytical investigation
of limiting cases (see Appendix) 
and by means of a linear stability analysis around the equilibrium state 
(in which we have $N_i(t) = N_i^0$, 
$Y_i(t) = Y_i^0$, $Q_i^0 - \sum_j a_{ij} Q_j^0 = Y_i^0$, and $P_i(t) = P_i^0$). 
Supply systems without a mechanism analogous to price adjustment are covered by 
$\alpha \rightarrow \infty$ and $f_i(P_i) = \mbox{const.}$ (no price sensitivity). 
\par
Denoting the $m$ eigenvalues of the input matrix $\mathbf{A}$ by $\omega_i$ with $|\omega_i| < 1$,
the $3m$  eigenvalues of the linearized model equations are $0$ ($m$ times) and 
\begin{equation}
 \lambda_{i,\pm} \approx \frac{1}{2} \bigg( - A_i \pm \sqrt{ (A_i)^2 - 4 B_i} \, \bigg) \, , 
\label{EV}
\end{equation}
where $A_i = \mu_i [C_i + \alpha_i D_i(1-\omega_i)]$, $B_i = \nu_i [C_i + \alpha_i D_i(1-\omega_i)]$,
$C_i = P_i^0 Y_i^0 |f'_i(P_i^0)| /N_i^0$ and $D_i = Q_i^0 /N_i^0$.
Formula (\ref{EV}) becomes exact when the matrix $\mathbf{A}$
is diagonal or the parameters $\mu_i C_i$, $\alpha_i \mu_i D_i$, $\nu_i C_i$ and $\alpha_i \nu_i D_i$ 
are sector-independent constants, otherwise the eigenvalues must be numerically determined. 
It turns out that the dynamic behavior
mainly depends on the parameters $\alpha_i$, $\nu_i/\mu_i{}^2$, and the eigenvalues $\omega_i$
of the input matrix $\mathbf{A}$ (see Fig.~1):  
In the case $\alpha_i \rightarrow 0$ 
of fast price adjustment, the eigenvalues  $\lambda_{i,\pm}$ are given by
$2\lambda_{i,\pm} = - \mu_iC_i \pm \sqrt{ (\mu_i C_i) ^2 - 4 \nu_i C_i}$, i.e.
the network structure does not matter at all. We expect
an exponential relaxation to the stationary equilibrium for $0 < \nu_i/ \mu_i{}^2 < C_i/4$, 
otherwise damped oscillations. Therefore, immediate price adjustments or similar mechanisms 
are an efficient way to stabilize economic and other supply systems. 
However, any delay ($\alpha_i > 0$) will cause damped or growing oscillations,
if complex eigenvalues  $\omega_i = \mbox{\rm Re}(\omega_i) + \mbox{\rm i\,Im}(\omega_i)$ exist.
\textit{Note that this is the generic case, as typical supply networks in natural and man-made systems are 
characterised by complex eigenvalues} (see top of Fig.~1). Damped oscillations can be shown to result
if all values $\nu_i/\mu_i{}^2 = \alpha_i \hat{\nu}_i/\hat{\mu}_i{}^2$ lie below the instability lines
\begin{eqnarray}
\nu_i/\mu_i{}^2 &\approx& \Big\{C_i + \alpha_i D_i [1-\mbox{Re}(\omega_i)]\Big\} \nonumber \\
&\times& \left( 1 + \frac{\{C_i+\alpha_i D_i [1-\mbox{Re}(\omega_i)]\}^2}
{[\alpha_i D_i\mbox{Im}(\omega_i)]^2} \right) 
\label{new}
\end{eqnarray}
given by  the condition $\mbox{Re}(\lambda_{i,\pm}) \le 0$. For identical parameters
$\nu_i/\mu_i{}^2 = \nu/\mu^2$ and $\alpha_i = \alpha$, the minimum of these lines agrees exactly 
with the numerically obtained curve in Fig.~1d. Values above this line cause 
small oscillations to grow over time, but the resulting aggregate behavior 
displays slow variations of small amplitude. The amplitude
is limited by phase shifts between the oscillations in different sectors and 
the non-linearities in Eqs.~(\ref{N}) to (\ref{Y}). 
Business cycles, for example, may therefore emerge without
exogenous shocks. This result markedly differs 
from the dominating business cycle theory \cite{econo}. 
It is surprising that increasing oscillation amplitudes are found if the adaptation rates $\nu_i$ are 
large. Nevertheless, many common production strategies suggest to keep constant inventories $N_i^0$, 
which potentially destabilizes economic systems. Ideal values of $\nu_i/\mu_i{}^2$ 
should lie below the instability line (\ref{new}), see Fig.~1d. 
\par
In some cases, all eigenvalues $\omega_i$ of the input matrix $\mathbf{A}$ are real.
This applies to symmetric matrices $\mathbf{A}$ and matrices equivalent to Jordan 
normal forms. Hence, the existence of loops in
supply networks is no sufficient condition for complex eigenvalues $\omega_i$ (see also Fig.~1f). It is also no
necessary condition. Examples for asymmetric matrices with real eigenvalues 
are linear (sequential) supply chains or regular distribution networks. 
In these cases, Eq.~(\ref{EV}) predicts a stable, overdamped behavior if 
all values $\nu_i/\mu_i{}^2  = \alpha_i \hat{\nu}_i/\hat{\mu}_i{}^2$ lie below the lines
\begin{equation}
 \nu_i/\mu_i{}^2 \approx  [ C_i + \alpha_i D_i (1-\omega_i)]/4 
\label{im}
\end{equation}
defined by $\min_i (A_i{}^2 - 4B_i) > 0$. For identical parameters
$\nu_i/\mu_i{}^2 = \nu/\mu^2$ and $\alpha_i = \alpha$, the minimum of these lines corresponds 
exactly to the numerically determined curve in Fig.~1h. Above it, one observes damped oscillations 
around the equilibrium state, but growing oscillations are not possible. 
In supply systems without a price adjustment or comparable mechanism,
Eq.~(\ref{im}) predicts an overdamped behavior for real
eigenvalues $\omega_i$ and $\hat{\nu}_i/\hat{\mu}_i{}^2 
< D_i(1-\omega_i)$ for all $i$, while Eq.~(\ref{new}) implies the stability condition
$\hat{\nu}_i/\hat{\mu}_i{}^2 
< D_i [1-\mbox{Re}(\omega_i)] \{1+ [1-\mbox{Re}(\omega_i)]^2/\mbox{Im}(\omega_i)^2\}$
for all $i$, given that some eigenvalues $\omega_i$ are complex.
\par
While previous studies have focussed on the synchronization of oscillators in different network topologies \cite{strogatz,PRL}, 
we have found that \textit{many supply networks display damped oscillations, even when their units---and
linear chains of these units---behave in an overdamped way. Furthermore, networks of damped oscillators 
tend to produce growing (and mostly asynchronous) oscillations.}
Due to the sensitivity of supply systems to their network structure,
network theory \cite{strogatz,nets,Stewart} 
can make useful contributions: On the basis of Eqs.~(\ref{new}) and (\ref{im}) one can 
design stable, robust, and adaptive supply networks (``network engineering''). 
For example, it is possible to identify structural and control policies which have
a dampening effect. However, in systems with competing goals 
(such as intersecting traffic streams), oscillatory solutions can be favourable.
The results presented in this study and the applied analytical techniques could be also
used and generalized to model the dynamics in metabolic networks \cite{Dane,metabo} 
to enhance the robustness of production processes, 
or to optimize disaster management \cite{disaster}. 

\section{Appendix: Mathematical Supplement}

{\em Linearized model equations:}
The linear stability analysis is based on the following
linearized equations for the deviations $n_i(t) = N_i(t) - N_i^0$, $p_i(t) = P_i(t) - P_i^0$,
and $q_i(t) = Q_i(t) - Q_i^0$ from the equilibrium state: 
\begin{eqnarray}
 \frac{dn_i}{dt} &=& q_i - \sum_j a_{ij} q_j - Y_i^0 f'_i(P_i^0) p_i - \xi_i(t) \, ,
\label{n} \\
 \frac{dp_i}{dt} 
 &=& \frac{P_i^0}{N_i^0} \left( - \nu_i n_i  - \mu_i \frac{dn_i}{dt} \right) \, ,
\label{p} \\
 \frac{dq_i}{dt} 
&=&  \frac{\alpha_i Q_i^0}{N_i^0} \left( - \nu_i n_i  - \mu_i  \frac{dn_i}{dt} \right) \, .
\label{q}
\end{eqnarray}
This system of coupled differential equations describes the response of the inventories,
prices, and production rates to variations $\xi_i(t)$ in the demand. The corresponding
eigenvalues are shown in Eq.~(\ref{EV}).
\par
{\em Dynamic behavior in limiting cases:}
Despite their mathematical similarity, Eqs.~(\ref{Q}) and (\ref{P}) have a surprisingly 
different impact on the macroeconomic dynamics: 
(i) In the case $\alpha_i \rightarrow 0$ of fast price adjustment,
one can eliminate Eq.~(\ref{q}) by assuming $q_i(t) \approx 0$ and $Q_i(t) \approx Q_i^0$, so that
$dn_i/dt \approx Y_i^0 |f'(P_i^0)| p_i(t) - \xi_i(t)$. Inserting (\ref{p}) into the time-derivative
of this equation finally results in the equations
\begin{equation}
\frac{d^2n_i}{dt^2} + \mu_i C_i \frac{dn_i}{dt}  + \nu_i C_i  n_i \approx - \frac{d\xi_i}{dt}
\end{equation}
of damped harmonic oscillators with eigenfrequencies $\omega_i^0 = \sqrt{\nu_i C_i}$, 
damping constants $\gamma_i = \mu_i C_i/2$, and
external driving $-d\xi_i/dt$ due to variations in the consumption rate. 
(ii) In the case $\alpha_i \gg 1$ of slow price adjustment or in supply networks for which
a price variable is not relevant, 
one can eliminate Eq.~(\ref{p}) by assuming $p_i(t) \approx 0$ and $P_i(t) \approx P_i^0$, so that
$dn_i/dt \approx q_i - \sum_j a_{ij} q_j - \xi_i$. Deriving this with respect to time and inserting
Eq.~(\ref{q}) delivers
\begin{equation} 
 \frac{d^2n_i}{dt^2} + \sum_j (\delta_{ij} - a_{ij}) \, \alpha_j D_j \left[ \mu_j \frac{dn_j}{dt} + \nu_j n_j(t) \right]
\approx - \frac{d\xi_i}{dt} \, , 
\end{equation}
where $\delta_{ij}=1$ for $i=j$, otherwise $\delta_{ij}=0$.  If we assume sector-independent
constants $\alpha_i \mu_i D_i = M$ and $\alpha_i \nu_i D_i = V$, 
the $2m$ eigenvalues $\lambda_{i,\pm}$ are given by
$2 \lambda_{i,\pm} = - M (1-\omega_i) \pm \sqrt{ [M(1-\omega_i)]^2 - 4 V (1-\omega_i) }.$
For empirical input matrixes $\mathbf{A}$, 
one never finds an overdamped, exponential relaxation to the economic equilibrium,  
but network-induced oscillations (see Fig.~1). An overdamped behavior is only possible
if all eigenvalues $\omega_i$ are real numbers. 
(iii) If $(A_i)^2/ B_i \gg 1$, the eigenvalues become $\lambda_{i,-} \approx - A_i$ and
$\lambda_{i,+} \approx - B_i / A_i = -\nu_i/\mu_i$. This corresponds
to a relaxation to the equilibrium state in the case of a
large responsiveness $\mu_i\gg 1$. An overdamped behavior is found if 
all eigenvalues $\omega_i$ are real numbers or if all $\alpha_i=0$, 
otherwise one expects network-induced oscillations.
Interestingly enough, $\mu_i \gg 1$ implies $(dP_i/dt)/(\mu_iP_i) \approx 0$, 
so that Eq.~(\ref{P}) reduces to $dN_i/dt \approx \nu_i [N_i^0 - N_i(t)]/\mu_i$. 
Therefore, $N_i(t)\approx N_i^0$ and $dQ_i/dt \approx 0$ (i.e. $Q_i\approx Q_i^0$).
Inserting this into Eq.~(\ref{N}) yields an implicit equation for the price $P_i(t)$ as a 
function of the fluctuations $\xi_i(t)$ in the consumption rate, 
as usually assumed in economics. It reads 
\begin{equation}
[Y_i^0 + \xi_i(t)] f_i(P_i(t)) \approx Q_i^0 - \sum_j a_{ij} Q_j^0 = \mbox{const.}
\end{equation} 
\par
{\em Boundary between damped and growing oscillations:} 
Starting with Eq.~(\ref{EV}),
stability requires the real parts $\mbox{Re}(\lambda_i)$ of all eigenvalues $\lambda_i$ to be non-positive. 
Therefore, the stability boundary is given by $\max_i \mbox{Re}(\lambda_i) = 0$. Writing 
$C_i + \alpha_i D_i(1-\omega_i) = \theta_i + \mbox{i} \beta_i$ with $C_i = P_i^0 Y_i^0 |f'_i(P_i^0)|/N_i^0$ and
defining 
\begin{eqnarray}
 \theta_i &=& C_i + \alpha_i D_i [1 - \mbox{Re}(\omega_i)] \, , \nonumber \\
 \beta_i &=& \mp \alpha_i D_i \mbox{Im}(\omega_i) \mbox{ (complex conjugate eigenvalues),} \nonumber \\
 \gamma_i &=& 4\nu_i/\mu_i{}^2 \, , 
\end{eqnarray}
we find 
\begin{equation}
 2\lambda_i/\mu_i 
= - \theta_i - \mbox{i} \beta_i + \sqrt{R_i+\mbox{i}I_i} 
\label{realim}
\end{equation}
with $R_i = \theta_i{}^2 - \beta_i{}^2 - \gamma_i\theta_i$ and $I_i=2\theta_i\beta_i - \gamma_i\beta_i$. 
The real part of (\ref{realim}) can be calculated via the relation
\begin{equation}
 \mbox{Re}\Big( \sqrt{R_i \pm \mbox{i}I_i}\Big) = \sqrt{\frac{1}{2} \Big( \sqrt{R_i{}^2 + I_i{}^2} + R_i\Big) } \, .
\end{equation}
The condition $\mbox{Re}(2\lambda_i/\mu_i) = 0$ is fulfilled by $\gamma_i = 0$ and
$\gamma_i = 4\theta_i (1 + \theta_i{}^2/\beta_i{}^2)$, i.e. the stable regime is given by
\begin{equation}
 \frac{\gamma_i}{4} = \frac{\nu_i}{\mu_i{}^2} = \frac{\alpha_i \hat{\nu}_i}{\hat{\mu}_i{}^2} \le 
 \theta_i \left( 1 + \frac{\theta_i{}^2}{\beta_i{}^2} \right) 
\end{equation}
for all $i$, corresponding to Eq.~(\ref{new}).
\par
{\em Boundary between damped oscillations and overdamped behavior:} For $\alpha_i > 0$, 
the imaginary parts of all eigenvalues $\lambda_i$ vanish if $\mbox{Im}(\omega_i)=0$
(i.e. $\beta_i =0$) and if $R_i \ge 0$. This requires 
\begin{equation}
\frac{4\nu_i}{\mu_i{}^2} = \gamma_i \le \theta_i - \frac{\beta_i{}^2}{\theta_i} 
= \theta_i = C_i + \alpha_i D_i (1-\omega_i)
\end{equation} 
for all $i$, corresponding to Eq.~(\ref{im}).

\clearpage
\begin{figure}[htbp]
\caption[]{Properties of our dynamic model of supply networks 
for a characteristic input matrix specified as average input matrix of macroeconomic commodity flows
of several countries (top) and for a synthetic input matrix generated by random changes of input matrix 
entries until the number of complex eigenvalues
was eventually reduced to zero (bottom). Subfigures
{(a), (e)} illustrate the color-coded input matrices $\mathbf{A}$, {(b), (f)} the
corresponding network structures, when only the strongest links (commodity flows) are
shown, {(c), (g)} the eigenvalues $\omega_i=\mbox{Re}(\omega_i) + \mbox{i} \,\mbox{Im}(\omega_i)$ 
of the respective input matrix $\mathbf{A}$, and
{(d), (h)} the phase diagrams indicating the stability behavior of the model equations (\ref{N}) to (\ref{Y})
on a double-logarithmic scale
as a function of the model parameters $\alpha_i = \alpha$ and $\nu_i/\mu_i{}^2  = \nu/\mu^2 = V/M^2$. The other model
parameters were set to $\nu_i = C_i = D_i = P_i^0 =N_i^0 = Y_i^0 = 1$.
Surprisingly, for empirical input matrices $\mathbf{A}$, 
one never finds an overdamped, exponential relaxation to the stationary equilibrium state, 
but network-induced oscillations due to complex eigenvalues $\omega_i$.} 
\label{Fig1}
\end{figure}
\end{document}